\documentclass[12pt]{article}
\usepackage{geometry}
\usepackage{a4}
\usepackage{graphicx}
\usepackage{epsf}
\usepackage{amsmath}
\usepackage{amssymb}
\usepackage{braket}
\usepackage{cite}
\newcommand{\be}{\begin{equation}}
\newcommand{\ee}{\end{equation}}

\newcommand{\Rmnum}[1]{\expandafter\@slowromancap\romannumeral #1@}
\newcommand{\bea}{\begin{eqnarray}}
\newcommand{\eea}{\end{eqnarray}}

\begin{document}
\def\A{{\mathbb{A}}}
\def\B{{\mathbb{B}}}
\def\C{{\mathbb{C}}}
\def\R{{\mathbb{R}}}
\def\s{{\mathbb{S}}}
\def\T{{\mathbb{T}}}
\def\Z{{\mathbb{Z}}}
\def\W{{\mathbb{W}}}
\begin{titlepage}
\title{Tidal Forces in Naked Singularity Backgrounds}
\author{}
\date{
Akash Goel, Reevu Maity, Pratim Roy, Tapobrata Sarkar
\thanks{\noindent E-mail:~ akagoel, reevu, proy, tapo @iitk.ac.in}
\vskip0.4cm
{\sl Department of Physics, \\
Indian Institute of Technology,\\
Kanpur 208016, \\
India}}
\maketitle
\abstract{
\noindent
The end stage of a gravitational collapse process can generically result in a black hole or a naked singularity. 
Here we undertake a comparative analysis of the nature of tidal forces in these backgrounds. The effect of such forces
is generically exemplified by the Roche limit, which predicts the distance within which a celestial object disintegrates due to the tidal effects of a 
second more massive object. In this paper, using Fermi normal coordinates, we numerically compute the Roche limit for a class of 
non-rotating naked singularity backgrounds, and compare them with known results for Schwarzschild black holes. Our analysis indicates 
that there might be substantially large deviations in the magnitudes of tidal forces in naked singularity backgrounds, compared to the 
black hole cases. If observationally established, these can prove to be an effective indicator of the nature of the singularity at a galactic centre. 
}
\end{titlepage}
\section{Introduction}
Einstein's theory of General Relativity (GR) \cite{Weinberg} is the most successful theory of gravity till date. An intriguing
prediction of the theory is the existence of mathematical singularities, which have been a major focus of research in gravity over the last century. Indeed, black holes, 
which are singular solutions of the Einstein's equations predicted by GR, form a basic ingredient in our understanding of the universe as it is generally 
believed that the centre of supermassive galaxies are candidate black holes. There are also other singular solutions of GR called naked singularities
that are not as well understood, but their existence cannot be ruled out. Possible distinctions between black holes and naked singularities
have been of great interest in the recent past. To substantiate and carry forward such analysis is the purpose of this paper. 

Recall that singularities in GR arise in gravitational collapse processes, see, e.g \cite{Poisson}. Indeed, In the absence of pressure, a spherically 
symmetric distribution of matter will collapse under its own gravity. This scenario is often repeated in more realistic situations including effects of pressure. 
The end stage of a generic collapse process is singular, and results in a black hole or a naked singularity \cite{Joshi1}, depending on initial conditions. 
While black holes are space-time singularities in GR that are covered
by an event horizon, naked singularities do not have such a cover. The cosmic censorship conjecture originally proposed
by Penrose more than four decades back states that nature does not allow naked singularities. However, a complete proof for this conjecture is lacking till date, and
several research groups have directed their attention to the formation of naked singularities, their stability, and in particular distinguishing features between
naked singularities and black holes. This last points assumes significance in predicting observational differences, if any, between black holes and naked 
singularities. Such analyses has been carried out for example in the context of accretion discs \cite{Joshi2} and gravitational lensing \cite{Virbhadra}, \cite{tapo1}.
In this work, we focus on another important aspect of gravity, namely tidal forces, and the purpose of this paper is to 
investigate the difference in the nature of these forces in naked singularity backgrounds, compared to black hole cases. The motivation for this line of
approach is that such differences might be significant to be observable. In particular, tidal effects on neutron stars \cite{Shapiro} in black hole backgrounds
have been extensively studied in the literature and one of our aims is to quantify the differences between these vis a vis naked singularity backgrounds, 
which might be relevant in realistic situations. 

Tidal forces are manifestations of non-local gravitational interactions. Consider, for example, a celestial object 
moving under the gravity of a more massive object. Non local gravity might cause this object to disintegrate. Viewed in the Newtonian context, this is easy to 
visualize. Consider for example a massive star of mass $M$ and radius $R$ that has a satellite made of incompressible matter, 
of mass $M_s$ and radius $r_s$, both objects assumed to be 
spherical, with their centres at a distance $d$ apart. An estimate of the tidal force can be obtained by computing the gravitational forces due to the star 
at the (near) end of the satellite and its centre. Assuming that $r_s/R \ll 1$ and equating this with the self gravitational force of the satellite predicts that 
the satellite will disintegrate if the dimensionless ratio  $(M_s/M)(d/R_s)^3$ is less than a critical value, this value depending on the spin angular velocity of 
the satellite. Then, this ratio translates into the fact that tidal disruptions occur if the radial separation between the star and the satellite is less than a 
critical value, known as the Roche limit. 

While tidal forces are easy to understand in the Newtonian context when the stellar mass is made of incompressible fluids, they may be substantially 
difficult to compute in the framework of GR, especially if one considers the fluid dynamics of the stellar object. The importance of this latter fact
has been recognized of late, and by now a large body of literature is available, which considers the nature of the fluid star that is being tidally disrupted.
Appropriate coordinates to deal with the problem in the framework of GR are the Fermi normal coordinates \cite{Manasse}. Here, one first 
sets up a locally flat system of coordinates along a geodesic, and then computes the tidal force taking into account the internal fluid dynamics
of the star, and hence obtain the Roche limit. The actual computation procedure is non-trivial, and with little analytical handle available, one usually
resorts to numerical analysis, along with a number of simplifying assumptions. 
This method was first elaborated upon in the work of Ishii et al \cite{ishii}, where tidal effects on objects in circular geodesics around a Kerr black hole 
were computed, upto fourth order in the Fermi normal coordinates. This built upon the work of Fishbone \cite{Fishbone}, and to the best of our knowledge, 
was the first GR computation of tidal effects taking into account the equilibrium fluid dynamics of
the stellar mass. The fluid star was treated in a Newtonian approximation, wherein it was possible to superpose a tidal potential calculated in Fermi normal
coordinates on the star's Newtonian potential, and a numerical routine was used to extract various physical quantities related to the tidal force. 
\begin{figure}[t!]
\centering
\includegraphics[scale=0.7]{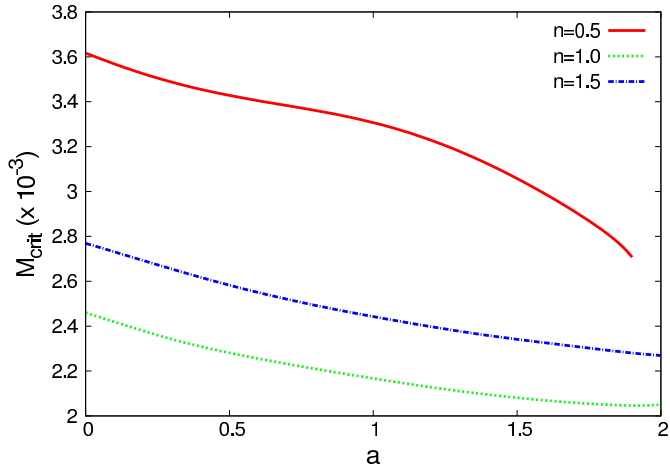}
\caption{Critical Mass For Circular geodesics in the Kerr geometry for different values of $a$, the rotation parameter. The singularity at $r=0$ is naked for $a>1$.}
\label{Kerr}
\end{figure}

The purpose of this paper is to generalize and extend the analysis of \cite{ishii} to naked singularity backgrounds. A simple possibility for example is to consider
tidal forces in the Kerr background considered in \cite{ishii}, and extend it to the regime where the singularity is naked. This might happen for example, when a black
hole accretes angular momentum by capturing rapidly spinning stellar objects. High spin black holes are known to exist in nature, and studies involving
the spin of the Kerr black hole lesser than, but close to its mass, have appeared in \cite{Liu}. Here we investigate the situation in which the spin of the black
hole exceeds its mass. 

We will, for most of this paper, adopt geometrized units in which the Newton's constant and the speed of light is set to unity. 
Then, the Kerr solution given in Boyer Lindquist
coordinates reads 
\begin{eqnarray}
ds^2 &=& -\left(1-\frac{2Mr}{\rho^2}\right)dt^2  + \frac{\rho^2}{\Delta}dr^2 + \frac{(r^2 + a^2)^2 - \Delta a^2{\rm sin}^2\theta}{\rho^2}{\rm sin}^2\theta d\phi^2 \nonumber\\
&+& \rho^2 d\theta^2  -\frac{4Mra{\rm sin}^2\theta}{\rho^2}dtd\phi~,
\end{eqnarray}
with $\rho^2 = r^2 + a^2{\rm cos}^2\theta$, and $\Delta = r^2 + a^2 - 2Mr$. Here, $M$ is the ADM mass and $a$, the angular momentum per unit mass of the source
is a parameter that denotes the spin of the black hole. 
For $a > M$, the singularity at $r=0$ is naked, and there is no event horizon. We will also set the ADM mass of the singularity, 
$M=1$ and the radius of the fluid star to $0.5$, a choice made consistently in all examples considered in this paper. 
Fig.(\ref{Kerr}) (obtained by methods illustrated in the later sections where we have used the Fermi normal 
coordinates upto second order) shows the variation of the critical mass below which a star in a circular geodesic orbit
is tidally disrupted, as a function of the rotation parameter (circular orbits in the Kerr naked singularity background and their stability has been studied extensively
in \cite{Stuchlik}. We use the result of that paper that all circular orbits in the Kerr naked singularity background are stable). Here we have chosen the radius of the
orbit of the star as $r=8$ (in geometrized units), and used a polytropic equation of state for the star (as given later in Eq.(\ref{poly})). 
The solid red, dot dashed blue and dotted green curves correspond to different values of the polytropic index ($0.5$, $1$, and $1.5$ respectively). 
The Kerr singularity is naked for $a>1$. We see from the figure that there is no special behavior of the critical mass near the region $a=1$, and that the
tidal force decreases with increasing angular momentum of the source. This latter fact implies that for the Kerr naked singularity background, 
tidal effects are smaller than those in Kerr black hole backgrounds. That this is not generically the case is one of the main results of this paper. 

Two things need to be kept in mind at this stage. First of all, we are using a vacuum solution of GR, which is an good approximation in the above example,
with $r=8$, i.e far from the central singularity. More realistic situations arise when there is diffused matter (possibly dark matter) around the central singularity, 
and this will be considered later in this paper. Secondly, as can be seen from Fig.(\ref{Kerr}), the change in the critical mass as a function of the rotation parameter 
is small. Hence, there is little hope to distinguish between a Kerr black hole and a Kerr naked singularity from the point of view of tidal disruptions. As we will see later in this
paper, other naked singularity backgrounds offer a better scenario, and that one could in principle obtain effects of tidal disruption that are orders of magnitude
apart from corresponding black hole situations. 

This paper is organized as follows. In the next section, we will first summarize the main ingredients used for our calculations and the various space-times
considered in this work. This involves three main steps. First we set up the Fermi normal coordinates for the naked singularity backgrounds. Next, the hydrostatic equations
for a fluid star at equilibrium are set up, and in the final step, a numerical routine is elaborated upon, for both circular and radial geodesic motions. 
The last two analyses discussed here essentially follow the work of Ishii et al \cite{ishii}. Towards the end of section 2, we will lay down all the approximations 
that we have made, and also comment on some of existing important results in the area. Then, 
in section 3, we present our main numerical results. Section 4 ends this paper with discussions and directions for future research.

\section{Computation of Tidal Effects in Naked Singularity Backgrounds}
We remind the reader that throughout this paper, we adopt geometrized units and set $G=c=1$. Units will be explicitly restored when we discuss some examples
towards the end of section 3. We will also use $T_{(ij)k}$ to denote the 
part of the tensor $T$ symmetric in indices $i$ and $j$. Latin and Greek indices are used to denote spatial and space-time coordinates respectively. To begin with,
we will illustrate the setup of the Fermi normal coordinates, following the work of  \cite{Manasse}. 

\subsection{Metric in Fermi Normal Coordinates}
In order to compute the tidal potential observed by a local inertial observer, we set up a locally flat system of coordinates on an arbitrary 
timelike geodesic, $G$. This would provide a way for a freely falling observer to report the effects of graviational field gradients in local experiments. 
We follow the standard procedure for constructing Fermi normal coordinates by setting up a tetrad basis at a point on the geodesic as the 
origin of our new coordinates, $\hat{e}^{\mu'}_{\nu}|_{P_0}	$ where unprimed indices denote the Fermi normal coordinates. We parallely 
transport the tetrad basis along $G$ and in particular choose $\boldsymbol{\hat{e}}_0(\tau)$ to be the tangent vector along $G$ which is, by 
definition, parallel transported. We hence obtain the tetrad basis on the entire geodesic as a function of the proper time. The Fermi normal 
coordinates, $x^{\alpha}$ of a point $P'$ in the neighbourhood of the geodesic are then specified. We define $x^0=\tau$ and $x^i$ corresponds to 
the point along the unique spacelike geodesic, $G'$ at proper distance $s$ with tangent vector at $P(\tau)$ given by direction cosines $x^i/s$ 
in the tetrad basis, $\boldsymbol{\hat{e}}_i$. 

In order to demonstrate the differences in tidal effects, we will consider general static, spherically symmetric spacetimes to cover a broad class 
of physical scenarios. We present an explicit computation of the Fermi normal coordinates along circular and radial geodesics for these. 
The class of metrics that we consider in this paper are given by 
\begin{equation}
\mbox{d}s^2=-A(r)\mbox{d}t^2+B(r)\mbox{d}r^2+C(r)\mbox{d}\Omega^2~,
\end{equation}
where the coefficients are positive functions of the radial coordinate and $\mbox{d}\Omega^2=\mbox{d}\theta^2+\sin^2\theta\mbox{d}\phi^2$ 
is the metric on the unit two sphere. 

The metric possesses two cyclic coordinates, $t$ and $\phi$ corresponding to the conserved quantities
\begin{equation}
E=A(r)\dot{t},~~~L=C(r)\dot{\phi}~,
\end{equation}
where $E$ and $L$ are the energy and angular momentum per unit mass of the test particle. 
From the remaining geodesic equations, we reduce the problem to the equivalent one-dimensional problem with effective potential
\begin{equation}
V(r)=\frac{1}{B(r)}\left[1+\frac{L^2}{C(r)}-\frac{E^2}{A{r}}\right]~.
\end{equation}
Imposing $V(r)=V'(r)=0$ for circular orbits gives
\begin{equation}
E=\frac{A(r)\sqrt{C'(r)}}{\sqrt{A(r)C'(r)-C(r)A'(r)}},~~~
L=\frac{ C(r)\sqrt{A'(r)}} {\sqrt{A(r)C'(r)-C(r)A'(r)}}~.
\end{equation}
We now set up the Fermi normal basis which must satisfy the parallel transport conditions
\begin{equation}
\boldsymbol{\hat{e}}_{\mu}.\boldsymbol{\hat{e}}_{\nu}=\eta_{\mu\nu},~~~
\nabla_{\alpha'}\left( \hat{e}^{\beta'}_{\mu } \right) \hat{e}^{\alpha'}_0=0~,
\label{eq:ptr}
\end{equation}
and $\boldsymbol{\hat{e}}_0$ is the tangent vector to the geodesic. This gives us the tetrad
\begin{eqnarray}
\hat{e}^{\alpha'}_0 &=&\left( \frac{E}{A(r)},0,0,\frac{L}{C(r)} \right) \nonumber\\
\hat{e}^{\alpha'}_1 &=&\left( -\frac{L\sin(\Omega\tau)}{\sqrt{A(r)C(r)}},\frac{\cos(\Omega\tau)}{\sqrt{B(r)}},0, -\frac{E\sin(\Omega\tau)}{\sqrt{A(r)C(r)}} \right) \nonumber\\
\hat{e}^{\alpha'}_2 &=&\left( 0,0,\frac{1}{\sqrt{C(r)}},0 \right)\nonumber\\
\hat{e}^{\alpha'}_3 &=&\left( \frac{L\cos(\Omega\tau)}{\sqrt{A(r)C(r)}},\frac{\sin(\Omega\tau)}{\sqrt{B(r)}},0, \frac{E\cos(\Omega\tau)}{\sqrt{A(r)C(r)}} \right)~.
\end{eqnarray}
Here, the constant angular velocity of motion is 
\begin{equation}
\Omega=D(r)\dot{\phi}=\frac{1}{2}\sqrt{\frac{A'(r)C'(r)}{A(r)B(r)C(r)}}~,
\end{equation}
where we have defined
\begin{equation}
D(r) =\frac{1}{2}\sqrt{\frac{C'(r)\left[A(r)C'(r)-A'(r)C(r)\right] }{A(r)B(r)C(r) }}~.
\end{equation}
In case of radial geodesics, proceeding similarly, we obtain
\begin{equation}
\dot{t}=\frac{E}{A(r)}~,~~~
\dot{r}=\sqrt{\frac{E^2-A(r)}{A(r)B(r)}}~.
\end{equation}
We set up the tetrad satisfying Eq.(\ref{eq:ptr}) using the geodesic equations in this case

\begin{eqnarray}
\hat{e}^{\alpha'}_0 &=&\left( \dot{t},\dot{r},0,0 \right) \nonumber\\
\hat{e}^{\alpha'}_1 &=&\left( \dot{r}\sqrt{\frac{B(r)}{A(r)}},\dot{t}\sqrt{\frac{A(r)}{B(r)}},0,0 \right) \nonumber\\
\hat{e}^{\alpha'}_2 &=&\left( 0,0,\frac{1}{\sqrt{C(r)}},0 \right) \nonumber\\
\hat{e}^{\alpha'}_3 &=&\left( 0,0,0,\frac{1}{\sqrt{C(r)}\sin\theta} \right)~.
\end{eqnarray}

\subsection{Calculating the Tidal Tensor}
Once the tetrad is constructed, we obtain the components of the curvature tensor in the tetrad basis
\begin{equation}
R_{\alpha\beta\gamma\delta}=\hat{e}^{\mu'}_{\alpha}\hat{e}^{\nu'}_{\beta}
\hat{e}^{\rho'}_{\gamma}\hat{e}^{\kappa'}_{\delta} R_{\mu'\nu'\rho'\kappa'}~.
\end{equation}
The metric of an observer in the Fermi normal coordinate system can be expanded as 
\begin{equation}
g_{\mu\nu}=\eta_{\mu\nu}+\frac{1}{2}g_{\mu\nu,ij}x^ix^j+\frac{1}{6}g_{\mu\nu,ijk}x^ix^jx^k+O(x^4)~.
\end{equation}
The expressions for the derivatives of the metric in the Fermi normal basis were computed in \cite{Manasse} and \cite{ishii}. 
The tidal potential, $\phi_t$ associated with the metric can then be expanded as 
\begin{equation}
\phi_t = -\frac{1}{2}(g_{00}+1)~ = \frac{1}{2}R_{0i0j}x^ix^j+\frac{1}{6}R_{0(i|0|j;k)}x^ix^jx^k+O(x^4)~.
\label{phit}
\end{equation}

In case of the circular geodesic, the non-vanishing derivatives of the metric needed for the computation of the tidal tensor at leading order are
\begin{eqnarray}
g_{00,11}&=&\frac{CA'^2C'-AC'^2A'-2ACC'A''+2ACA'C''}{4ABC(AC'-A'C)}\nonumber\\
g_{00,22}&=&\frac{A'}{A'C-AC'},~~g_{00,33} = -\frac{A'C'}{4ABC}~.
\end{eqnarray}

For radial geodesic, the energy of the test particle $E$ is a free parameter.  This parameter cannot however assume large values, as we 
discuss in subsection 2.5. The corresponding non-zero quantities in this case are 
\begin{eqnarray}
g_{00,11}&=&\frac{A^2[-2ABA''+A'(BA'+AB')]}{4A^4B^2}\nonumber\\
g_{00,22}&=&g_{00,33} = -\frac{E^2BCA'C'+A(E^2-1)(BC'^2+C(B'C'-2BC''))}{4A^2B^2C^2}~.
\end{eqnarray}

In our analysis, we will focus on the following class of static, spherically symmetric metrics to compare the black hole and naked singularity cases-
\begin{itemize}
\item {\em Janis Newmann Winicour (JNW) spacetime} \cite{JNW1},\cite{JNW2}- This is a solution to the Einstein-Klein-Gordon system of 
equations. The spacetime consists of a singularity that is globally naked at $r=B$.
\begin{equation}
\mbox{d}s^2=-\left(1-\frac{B}{r}\right)^{\nu}\mbox{d}t^2+\left(1-\frac{B}{r}\right)^{-\nu}\mbox{d}r^2+r^2\left(1-\frac{B}{r}\right)^{1-\nu}\mbox{d}\Omega^2~.
\label{JNWmetric}
\end{equation}
This spacetime is sourced by the scalar field with magnitude $q$
\begin{equation}
\psi=\frac{q}{B\sqrt{4\pi}}\ln\left(1-\frac{B}{r}\right)~.
\end{equation}
Here, $B$ is a parameter related to the ADM mass by $B=2\sqrt{q^2+M^2}$ and $\nu=2M/B$ lies between 0 and 1.
Note that the case $\nu=1$ corresponds to the Schwarzschild metric which is a black hole solution. For a fixed value of the ADM mass, 
moving away from $\nu = 1$ takes us
deep into the naked singularity regime. Note however that there is a tradeoff here, namely that smaller values of $\nu$ implies larger values of 
$B$, which in turn means that the solution is valid from a higher value of the radial coordinate $r$. Also, we would be mostly interested in 
stable circular orbits, and the criteria for such orbits are well established. We simply quote the result that defining the variable 
\begin{equation}
r_{\pm} = \frac{B}{2}\left( 1 + 3\nu \pm \sqrt{5\nu^2 - 1}\right)~,
\end{equation}
it can be checked that for $0 < \nu < 1/\sqrt{5}$, stable circular orbits exits for all values of the radius (greater than $B$). For $\nu$ lying between
$1/\sqrt{5}$ and $1/2$, stable circular orbits are possible for radii less than $r_-$ and greater than $r_+$, while for $\nu$ greater than $1/2$, such
orbits are possible for all values of the radius greater than $r_+$. In our analysis on the JNW spacetimes, we have chosen values of the radii that
correspond to stable circular orbits. 
\item {\em Bertrand spacetimes (BSTs)} \cite{Perlick},\cite{dbs2}- These are seeded by matter that can be given an effective two-fluid description and are 
a candidate for galactic dark matter. We look at the type II BST
\begin{equation}
\mbox{d}s^2=-\frac{1}{D+\frac{\alpha}{r}}\mbox{d}t^2+\frac{1}{\beta^2}\mbox{d}r^2+r^2\mbox{d}\Omega^2~.
\label{BSTmetric}
\end{equation}
Here, $\beta$ is a rational. The parameters $\alpha$ and $\beta$ can be used to provide a phenomenological estimate for the total dark matter mass 
of the galaxy (in good agreement of known results for low surface brightness galaxies) as \cite{dbs1}
\begin{equation}
M=\frac{\alpha R_g^2}{2(\alpha+DR_g)^2}~,
\label{massBST}
\end{equation}
where a good estimate for the size of the galaxy is $R_g=\alpha/D$. It can be checked that here we have a central singularity at $r=0$, which is naked. 
For computations performed with BSTs, we have set $M=1$, which implies, via Eq.(\ref{massBST}) that $\alpha = 8D^2$. This will be understood
in sequel. It can be checked that for BSTs, all circular orbits are stable. 
\item {\em Joshi Malafarina Narayan (JMN) spacetime} \cite{JMN1},\cite{JMN2} - describes a class of geometries that result from the 
gravitational collapse of a matter cloud with regular initial conditions into an astympotically static equilibrium configuration containing a central naked singularity at $r=0$
\begin{equation}
\mbox{d}s^2=-(1-M_0)\left(\frac{r}{R_0}\right)^{\frac{M_0}{1-M_0}}\mbox{d}t^2+\frac{1}{1-M_0}\mbox{d}r^2+r^2\mbox{d}\Omega^2~.
\label{JMNmetric}
\end{equation}
This metric also matches smoothly to a Schwarzschild exterior at $r=R_0$ with mass $M=M_0R_0/2$. The system possesses stable circular orbits for $M_0\leq 2/3$. 
\end{itemize}
While the JNW, BST and JMN space-times discussed above contain naked singularities, towards the end of this paper, we will also briefly comment 
on the nature of tidal forces in the interior Schwarzschild solution. This describes the metric of a fluid of constant density that is matched with an external
Schwarzschild solution of mass $M$, and reads
\begin{equation}
\mbox{d}s^2=-\left( \frac{3}{2}\sqrt{1-\frac{2M}{R_0}} - \frac{1}{2}\sqrt{1-\frac{2Mr^2}{R_0^3}} \right)^2\mbox{d}t^2+
\left(1-\frac{2Mr^2}{R_0^3}\right)^{-1}\mbox{d}r^2+r^2\mbox{d}\Omega^2~.
\label{IntSchmetric}
\end{equation}
The matching radius here is $r=R_0$.

\subsection{Hydrostatic equations for Computing Equilibrium Configurations}
We will now review a formulation to compute the equilibrium configurations under gravitational tidal forces arising from the potential 
gradients in a curved spacetime geometry. The discussion in this and the next subsection essentially follows \cite{ishii}, and we closely follow the
notation of that paper.
We assume that the radius of the star \footnote{By ``radius'' of a stellar object, we really mean the length of its major axis. This slight abuse of notation should be kept in
mind.} is much smaller than the radius of its orbit so that in Fermi normal coordinates, the self gravity of the star 
is described by Newtonian gravity, with an additional tidal potential due to the curved background. In such a case, the fluid star obeys the hydrodynamic equation
\begin{equation} \label{euler}
\rho\frac{\partial v_i}{\partial \tau} + \rho v^j \frac{\partial v_i}{\partial x^j}=-\frac{\partial P}{\partial x^i}-
\rho\frac{\partial (\phi+\phi_{t})}{\partial x^i}+\rho\left[ v_j\left( \frac{\partial A_j}{\partial x^i} -\frac{\partial A_i}
{\partial x^j} \right) -\frac{\partial A_j}{\partial \tau}  \right]~.
\end{equation}
This equation is analogous to the Euler equation with the tidal potential $\phi_{t}$ superposed with the Newtonian potential $\phi$ and an additional term 
associated with the gravitomagnetic force described by the vector potential $A_k=\frac{2}{3}R_{k(ij)0}x^ix^j$.
Here, $v^i=dx^i/d\tau$ is the fluid three-velocity and $P$ is the fluid pressure. The Newtonian potential produced due to the mass of the star obeys the Poisson equation
$\Delta \phi=4\pi\rho$.

As is common in the literature, we assume a polytropic equation of state for the star given by
\begin{equation}
P=\kappa \rho^{1+\frac{1}{n}}~,
\label{poly}
\end{equation}
where $\kappa$ and $n$ are polytropic constants. The equation of continuity, along with that for hydrostatic equilibrium then leads to the
Lane-Emden equation
\begin{equation}
\frac{1}{\xi^2}\frac{d}{d\xi}\left( \xi^2\frac{d\theta}{d\xi} \right)+\theta^n=0~,
\end{equation}
where $\xi$ is a dimensionless radial parameter obtained by scaling the radius, and $\rho=\rho_0\theta^n$, where
$\rho_c$ is the central density. In our computations, this will be used to fix the stellar radius in terms of the Lane-Emden coordinate 
at the stellar surface $\xi_0$
\begin{equation}
R=\left(\ \frac{(n+1)\kappa\rho_c^{(1-n)/n}}{4\pi} \right)^{1/2}\xi_0~.
\end{equation}
We will consider the values of the polytropic index $n=$ $0.5$, $1$ and $1.5$ for which it can be checked that $\xi_0 =$ $2.75$, $\pi$ and $3.65$ respectively.
 
We first consider the case of circular geodesics. Following \cite{ishii}, we assume a co-rotational velocity field for the fluid star with the velocity field 
\begin{equation}
v^i=\left[ -{x^3-x_c\sin(\Omega\tau)},0,{x^1-x_c\cos(\Omega\tau)} \right]~,
\end{equation}
where $x_c$ is a small correction constant. 
In order to simplify numerical computations, we go to a frame where the Euler equation is independent of $\tau$. 
This is achieved by choosing the frame, $\tilde{x}^i$ rotating at angular velocity $\Omega$ about $x^2$ axis. 
The hydrostatic equation corresponding to the first integral of the Euler equation in this frame is then given as
\begin{equation}\label{int}
\frac{\Omega^2}{2}\left[(\tilde{x}^1-2x_c)^2+(\tilde{x}^3)^2\right] =  \kappa(n+1)\rho^{\frac{1}{n}} + \phi +\phi_{t} + \phi_{m} + C~.
\end{equation}
Here $\phi_m$ is the contribution due to the gravitomagnetic force and we have used the polytropic equation. 

In case of radial geodesics, \footnote{There is an assumption of instantaneous equilibrium here. See the discussion in subsection 2.5.} 
we take the velocity field to be irrotational, and hence the gradient of a velocity potential,
\begin{equation}
v_i+A_i=\frac{\partial\psi}{\partial x^i}~.
\end{equation}
The first integral for \eqref{euler} is then given as
\begin{equation} \label{eulerint}
-\frac{\partial\psi}{\partial\tau}-\frac{1}{2}\delta_{ij}\frac{\partial\psi}{\partial x^i}\frac{\partial\psi}{\partial x^j}= 
\kappa(n+1)\rho^{\frac{1}{n}} + \phi +\phi_{t} -\frac{\partial\psi}{\partial x^j}A^j + C~.
\end{equation}
Here, $\psi$ is determined by solving the equation of continuity,
\begin{equation} \label{continuity}
\rho\Delta\psi+\delta_{ij}\frac{\partial\psi}{\partial x^i}\frac{\partial\rho}{\partial x^j}=0~.
\end{equation}
These, along with the Poisson equation are the governing equations for the radial case. 

\subsection{Numerical Procedure}
For our numerical procedure, we again closely follow \cite{ishii}. 
We first consider the circular case. Here the governing equations are the Poisson equation and Eq.\eqref{int}, and these are to be solved iteratively. 
In order to achieve a convergence in the iteration, we switch to a dimensionless coordinate, $\tilde{x}^i= h q^i$. We correspondingly 
define the rescaled potentials, $\bar{\phi}=\phi/h^2$, $\bar{\phi_t}=\phi_t/h^2$ so that our system of equations reduces to
\begin{equation}
\Delta_h \bar{\phi}=4\pi\rho~,
\end{equation}
\begin{equation}\label{euler2}
\frac{\Omega^2}{2}h^2\left[(q^1-2q_c)^2+(q^3)^2\right] =  \kappa(n+1)\rho^{\frac{1}{n}} + h^2(\bar{\phi} +\bar{\phi}_{t} + \bar{\phi}_{m}) + C~.
\end{equation}
This system contains three primary constants, namely $h$, $q_c$ and $C$. The remaining constants are fixed by choosing units so that the Newtonian 
mass of the central body is unity or by choosing the metric parameters in these units. The equilibrium configurations will be computed for 
different values of $\rho_c$. The three primary constants will have to be fixed at each step of the iteration. 
As in \cite{ishii}, we solve the Poisson equation in Cartesian coordinates on a uniform grid of size (2$N$+1, $N$+1, 2$N$+1) in order to cover the 
region $-L\leq q^1\leq L$, $0\leq q^2\leq L$, $-L\leq q^3\leq L$. Here, we have assumed reflection symmetry with respect to the $q^2=0$ plane. 
We typically choose $N=50$ and the grid spacing to be $q_s/40$, where $q_s$ is the coordinate at the stellar surface along the $q^1$ axis. 
We use a Fortran subroutine to solve the fourth-order finite difference 
approximations to the elliptic partial difference equations. We implement a numerical algorithm that is summarized as :
\begin{itemize}
\item We take a trial density distribution and use the cubic grid-based Poisson solver to solve the Poisson equation. We impose Neumann 
boundary conditions on the $q^2=0$ plane to achieve reflection symmetry along the $q^2$ axis and Dirichlet boundary conditions elsewhere, namely,
\begin{equation}
\phi\to -\frac{1}{r}\int \rho\,d^3r + O(r^{-3})~.
\end{equation}
The distribution $\phi$ is obtained which is used at the next step. 
\item The constants $h$, $q_c$, $C$ are determined from the Euler equation by imposing constraints on the density profile. We match the 
central density, $\rho(0,0,0)=\rho_c$ and require that $\partial\rho/\partial q^1=0$ at the center. Also, we require the density to 
vanish at the stellar surface, $\rho(q_s,0,0)=0$. The set of constraints when substituted in \eqref{euler2} can be solved to determine the 
values of the constants. From these values, the new distribution $\rho(q^i)$ is computed again using \eqref{euler2}. 
\item The new density distribution is truncated at the first instance where it goes to zero and is again used as a source in the Poisson equation. 
The process is repeated iteratively until it converges to the equilibrium distribution and the density profile becomes stationary upto a numerical tolerance value.
\item We compute several equilibrium configurations for decreasing $\rho_c$ and track the quantity $\chi=\partial \rho/\partial q^1$ at the 
stellar surface ($q_s$, 0, 0). The critical value of $\rho_c$ corresponding to the tidal disruption limit i.e. the Roche limit is obtained 
when $\chi$ goes abruptly from a positive value to zero. Beyond this point, the density distribution becomes flat, signalling tidal disruption. 
For positive $\chi$, the star is in a stable configuration. 
\end{itemize}

We can adopt a similar procedure for radial geodesics where there is an extra function $\psi$ that has to be determined. This can be
 handled using the additional constraint imposed by the continuity equation. We proceed as follows. As before, a trial density function is 
 used to solve the Poisson equation and obtain $\phi$. This initial density distribution is also used to solve the elliptic partial differential Eq.\eqref{continuity} 
 using a three dimensional Cartesian PDE solver based on multigrid iteration. Once $\phi$ and $\psi$ are known, we use Eq.\eqref{eulerint} to 
 determine the contants $h$ and $C$. This is done by imposing two constraints, namely $\rho=\rho_c$ at the center and the density must vanish at 
 the stellar surface. One can now use the values of these cosnstants in Eq.\eqref{eulerint} to determine $\rho$ on the entire grid. This new density 
 distribution in again fed into the Poisson and continuity equations and the iterative procedure is used as previously to compute the equilibrium 
 configuration, and hence the Roche limit. 

\subsection{Approximations, Limitations and Related Issues}

Before we move on to present our results, we point out the approximations that we will make, and the limitations of our analysis and possible caveats that
need to be kept in mind. 

Since we are mainly interested in demonstrating the difference in tidal forces between naked singularities and black hole backgrounds, it suffices to work up to 
second order in the tidal potential. Fourth order effects can be included in our analysis, but these, apart from being small, will not qualitatively change our results. 
A more refined analysis of the scenarios presented here including higher order corrections to the metric in Fermi normal
coordinates is left for the future. Further, the contribution of the gravitomagnetic term in the hydrodynamic equations is only significant at fourth order in 
the expansion and so, these terms can be neglected at our level of approximation. Also, for the radial geodesics, for simplicity, we have chosen the velocity 
potential to be zero to give the irrotational field. In the rest of the paper, we will proceed with these approximations. 

There are a few caveats that we need to keep in mind. First, note that we work in a probe approximation,
where the effects of the star back reacting on the metric is ignored. Effects of back-reaction are somewhat intractable in the present formalism and pose
a formidable challenge. Here we proceed with the assumption that the background metric is fixed.

Second, in case of radial geodesics, our analysis for the Roche radius is valid under the approximation of instantaneous equilibrium. 
In general, this radius will be less (or equivalently, the critical mass will be smaller)  than that calculated from the numerical simulation by a small 
correction  equal to the distance travelled by the star in the delay period during which the star attains the equilibrium configuration. The approximation 
is valid in the limit that the distance travelled in the time scale during which the star attains equilibrium is smaller than the distance over which the tidal force 
changes appreciably. This condition is most naturally satisfied for low energy scales. We will, for radial geodesics, set $E = 10^{-2}$ and it is assumed that at 
our level of approximation, the tidal force is constant over the range of validity of the Fermi normal expansion.

Next, we note that a general tidal model can also result in a weak tidal encounter between a star orbiting in a massive background. 
Such encounters can result in a mass loss from the star, as discussed in details in \cite{cheng}. These effects on white dwarfs are seen to 
increase the likelihood of tidal disruption as a function of time. In this work, we will however restrict ourselves to conditions for complete tidal 
stripping for both radial and circular orbital motion.

Finally, although not presented here, the present analysis can be used to build upon the work of \cite{kesden} to calculate rates for tidal 
disruption events for supermassive black holes capturing not only the relativistic treatment of tidal disruption but also the hydrodynamics of the orbiting star. 
The method is summarized as follows. Monte Carlo simulations are performed with an appropriate distribution of the free variables in the geodesic equations 
for a general orbit. These equations, in Fermi normal coordinates, are integrated upto the pericenter. In our case, we can run the numerical code to determine if the star 
is directly captured or is tidally disrupted at this point by computing the equilibrium hydrodynamic configuration. This should give us an estimate for the tidal 
disruption rate of the simulated orbits. One can then study the effects of varying different metric parameters and astronomical data from flares associated 
with these tidal disruption events can be used to provide insight into the nature of the background singularity. In principle, this should improve upon the work 
of \cite{kesden}, but such an analysis is left for the future. 

\section{Results and Analysis}

In this section, we will present the results of our numerical analysis. We remind the reader that unless mentioned otherwise, 
in all examples below, we have set the radius of the star to be $0.5$, and the mass of the background to unity. 
We start with circular geodesics in various space-times. 
\begin{figure}[t!]
\begin{minipage}[b]{0.5\linewidth}
\centering
\includegraphics[scale=0.6]{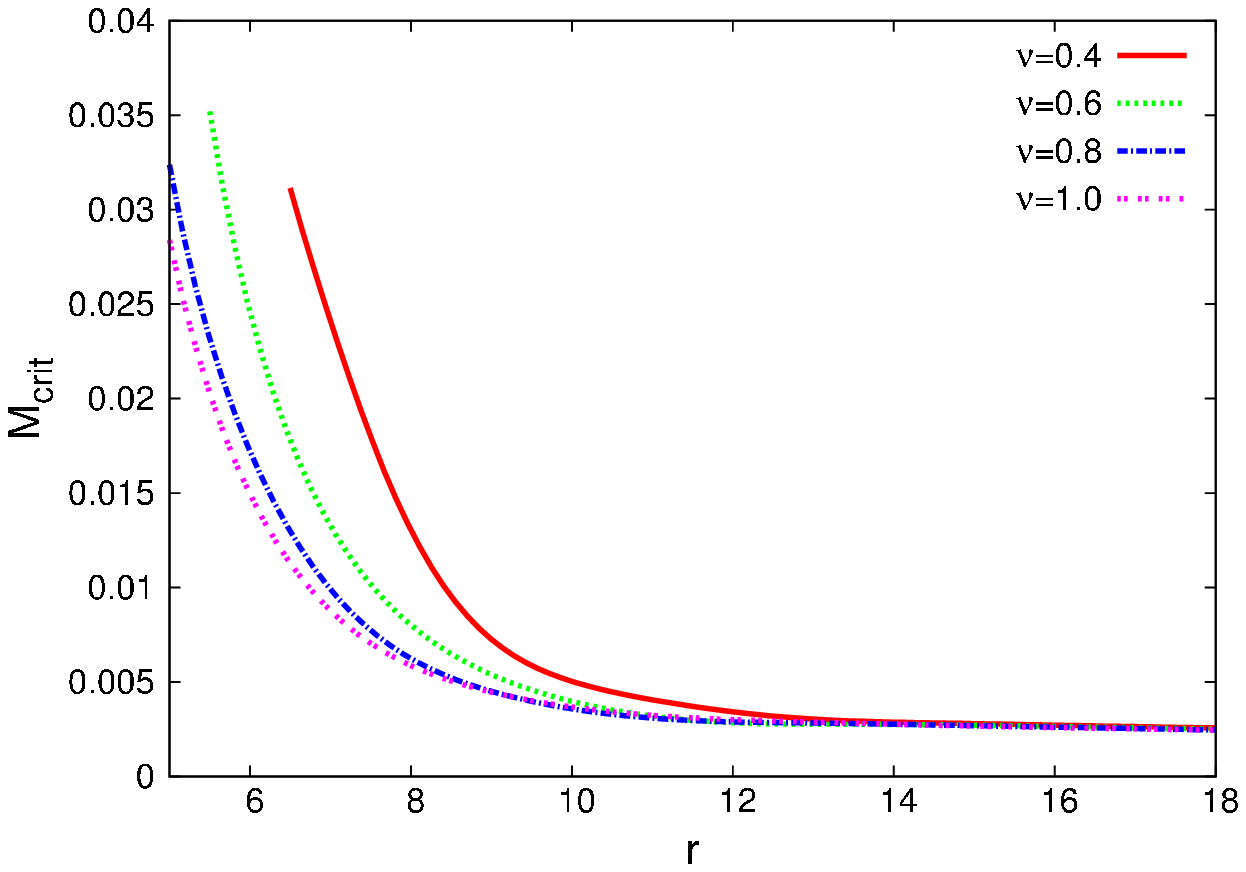}
\caption{Critical Mass for circular geodesics in JNW space-times for different $\nu$, with $n=0.5$}
\label{jnwcirc}
\end{minipage}
\hspace{0.2cm}
\begin{minipage}[b]{0.5\linewidth}
\centering
\includegraphics[scale=0.6]{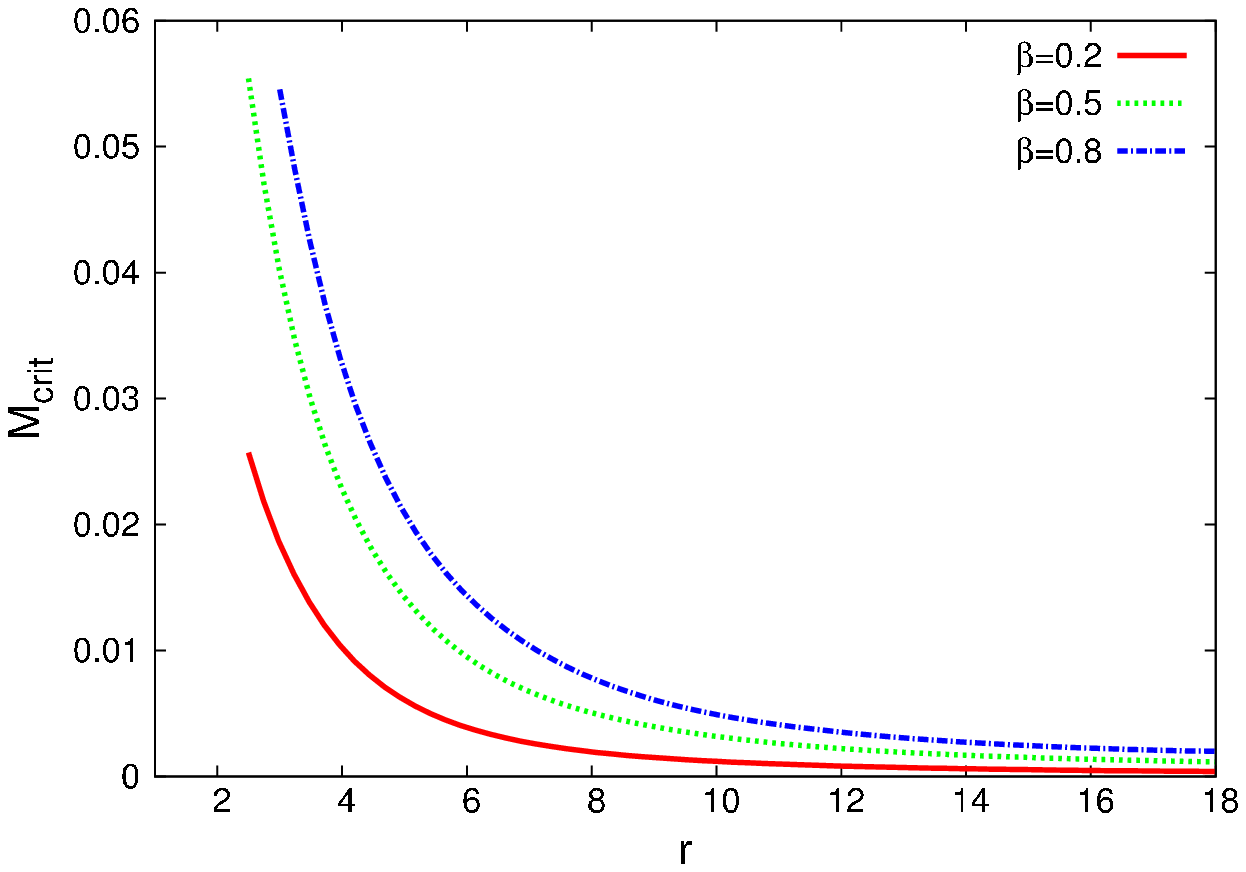}
\caption{Critical Mass for circular geodesics in BSTs for different $\beta$, with $D=10^5$, $n=0.5$}
\label{bstcirc}
\end{minipage}
\end{figure}
In Fig.(\ref{jnwcirc}), we show the critical mass as a function of the radial distance for JNW space-times, with the polytropic index (of Eq.(\ref{poly}))
$n=0.5$. The solid red, dotted green and dot-dashed blue curves correspond to the values $\nu =$ $0.01$, $0.1$ and $0.6$ respectively,
in Eq.(\ref{JNWmetric}). For comparison, we have shown with the dashed black curve the corresponding result for the Schwarzschild black hole
($\nu = 1$ in Eq.(\ref{JNWmetric})). Clearly, the effect of this naked singularity background is seen to increase the tidal force on the star. 
This is our first observation : if, at a given radius, stellar objects above the critical mass predicted from the Schwarzschild black hole are seen to exist, they might
point towards a naked singularity, rather than one with a horizon. 

In Fig.(\ref{bstcirc}), the computation is repeated for the BST naked singularity background of Eq.(\ref{BSTmetric}). In this figure, the solid
red, the dotted green and the dot-dashed blue curves correspond to choosing the values $\beta$ of Eq.(\ref{BSTmetric}) as $0.2$, $0.5$ and $0.8$, 
respectively. We see here that increase in the value of $\beta$ corresponds to higher tidal forces. This is again indicative of the fact that for 
similar central masses, the BST naked singularity predicts higher tidal disruption limits. If observational indications of this fact are found in future
experiments, BSTs can possibly be used as models to understand such a scenario.  

\begin{figure}[t!]
\begin{minipage}[b]{0.5\linewidth}
\centering
\includegraphics[scale=0.6]{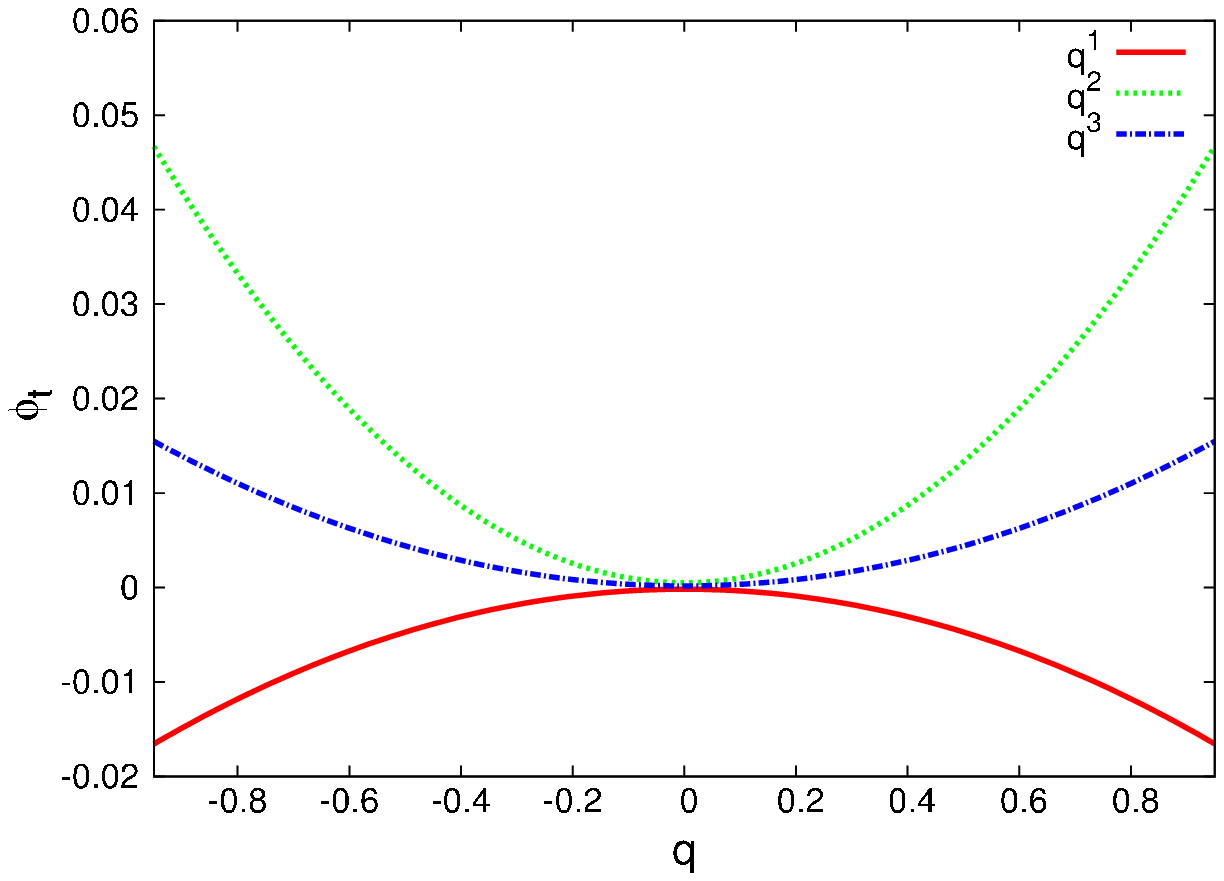}
\caption{Leading Order Tidal Potential for Circular BST, with the chosen values $D=10^5$, $\beta=0.8$, $r=3$.}
\label{bsttidpot2}
\end{minipage}
\hspace{0.2cm}
\begin{minipage}[b]{0.5\linewidth}
\centering
\includegraphics[scale=0.6]{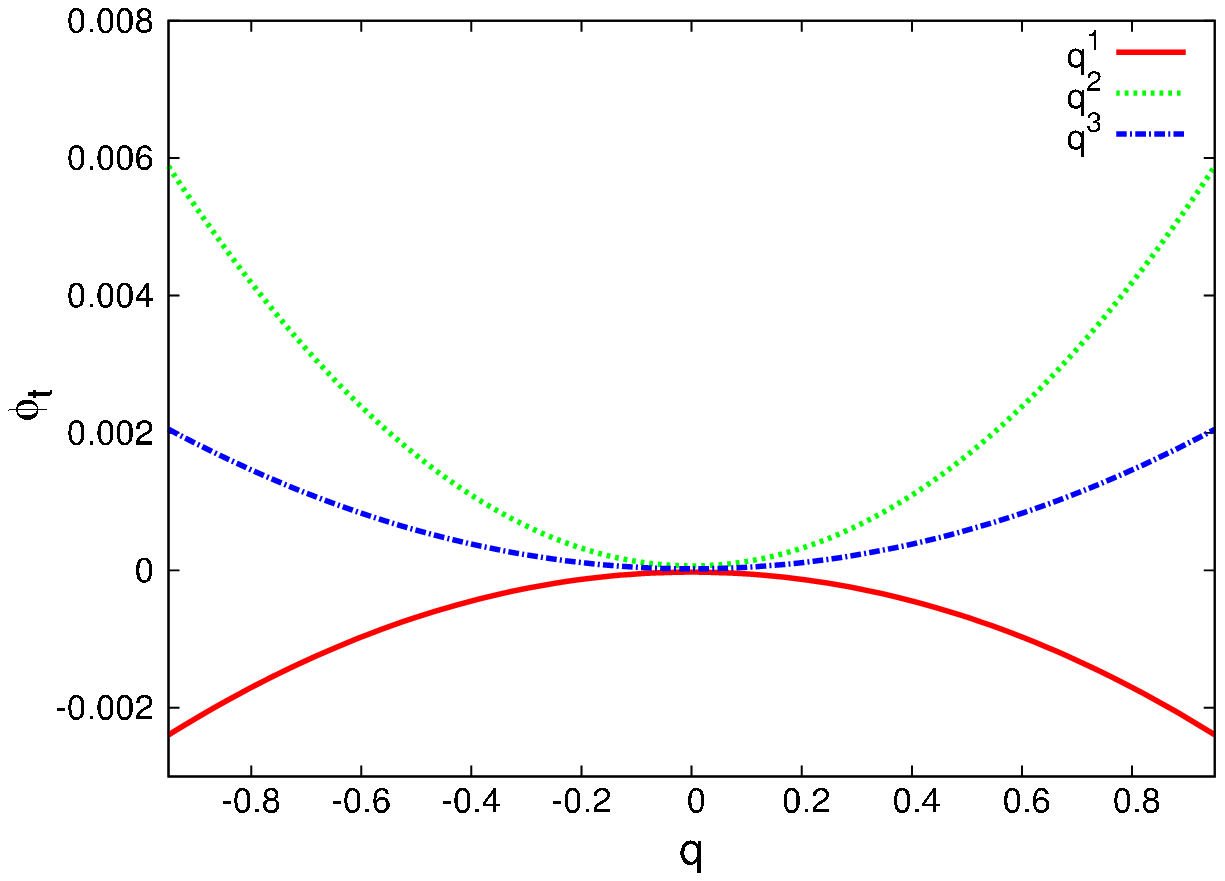}
\caption{Leading Order Tidal Potential for Circular BST, with the chosen values $D=10^5$, $\beta=0.8$, $r=8$.}
\label{bsttidpot}
\end{minipage}
\end{figure}
\begin{figure}[t!]
\begin{minipage}[b]{0.5\linewidth}
\centering
\includegraphics[scale=0.6]{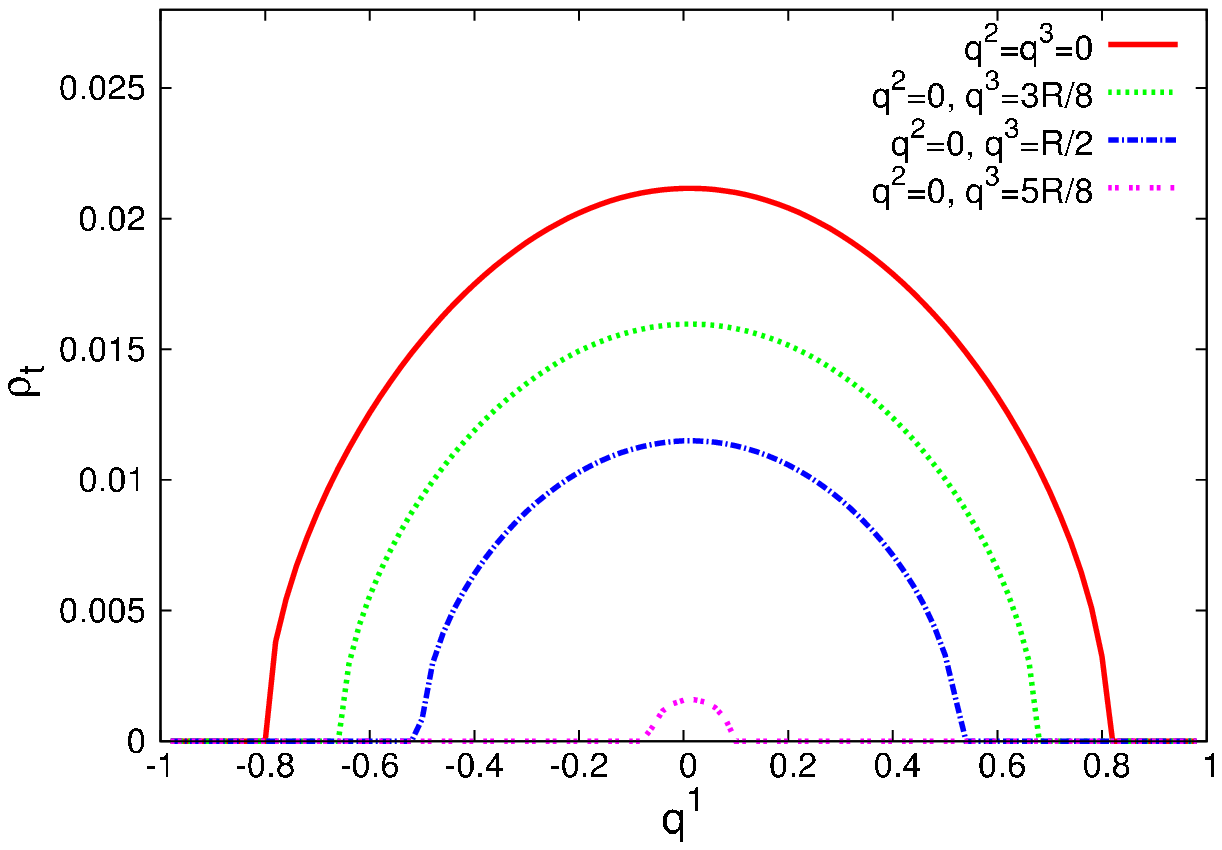}
\caption{Density variation near equilibrium for BST (Circular Orbit), with $D=10^5$, $\beta=0.8$, $n=0.5$}
\label{bstden1}
\end{minipage}
\hspace{0.2cm}
\begin{minipage}[b]{0.5\linewidth}
\centering
\includegraphics[scale=0.6]{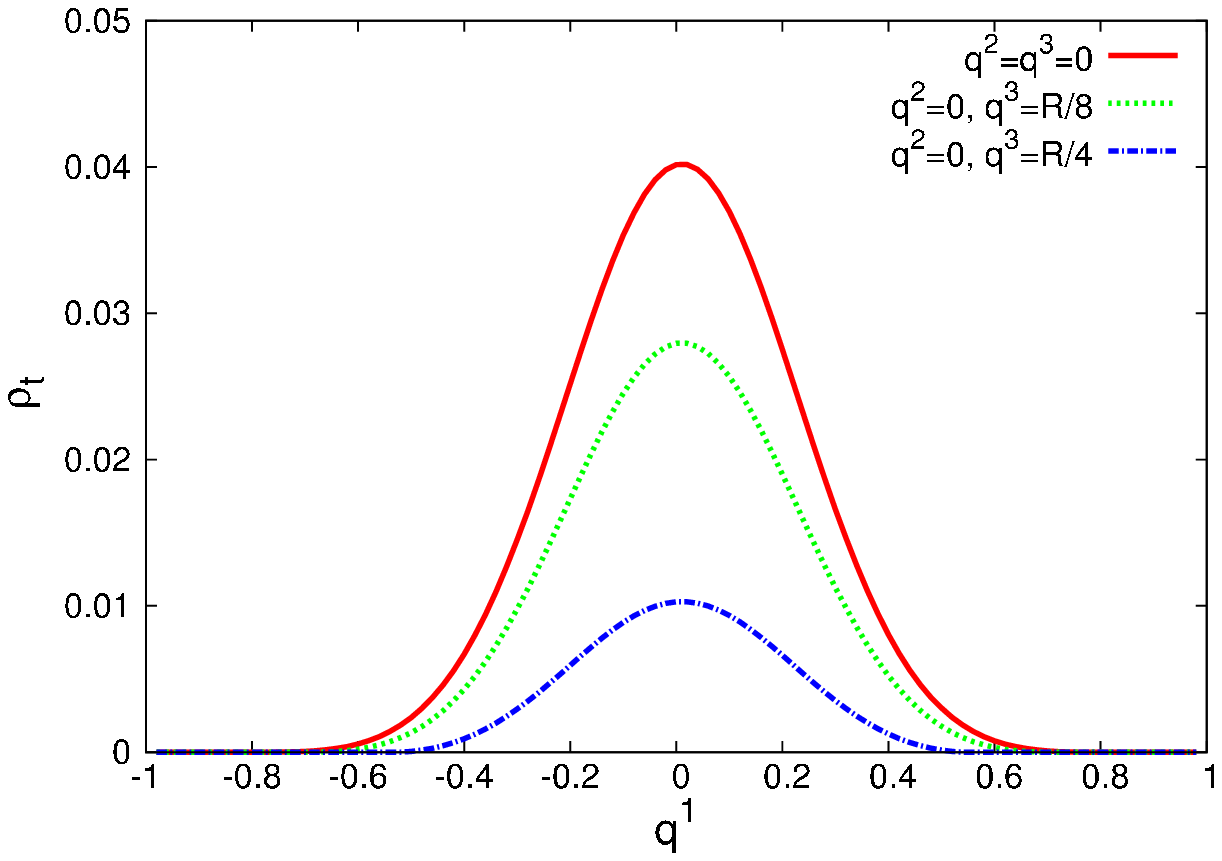}
\caption{Density variation near equilibrium for BST (Circular Orbit), with $D=10^5$, $\beta=0.8$, $n=1.5$}
\label{bstden2}
\end{minipage}
\end{figure}
\begin{figure}[t!]
\begin{minipage}[b]{0.5\linewidth}
\centering
\includegraphics[scale=0.6]{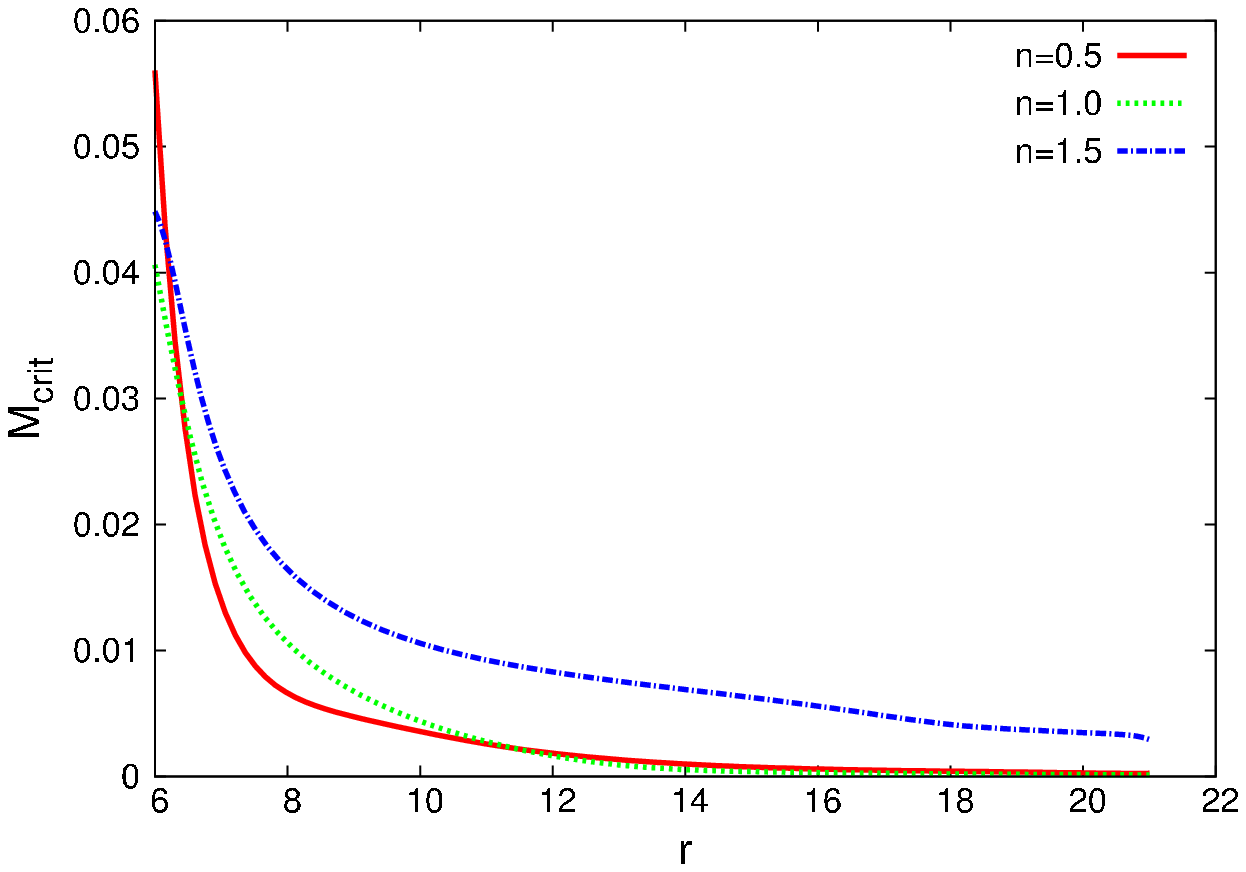}
\caption{Critical mass for radial geodesics in the JNW background for different polytropic index, where we have taken $\nu = 0.4$.}
\label{radjnwn}
\end{minipage}
\hspace{0.2cm}
\begin{minipage}[b]{0.5\linewidth}
\centering
\includegraphics[scale=0.6]{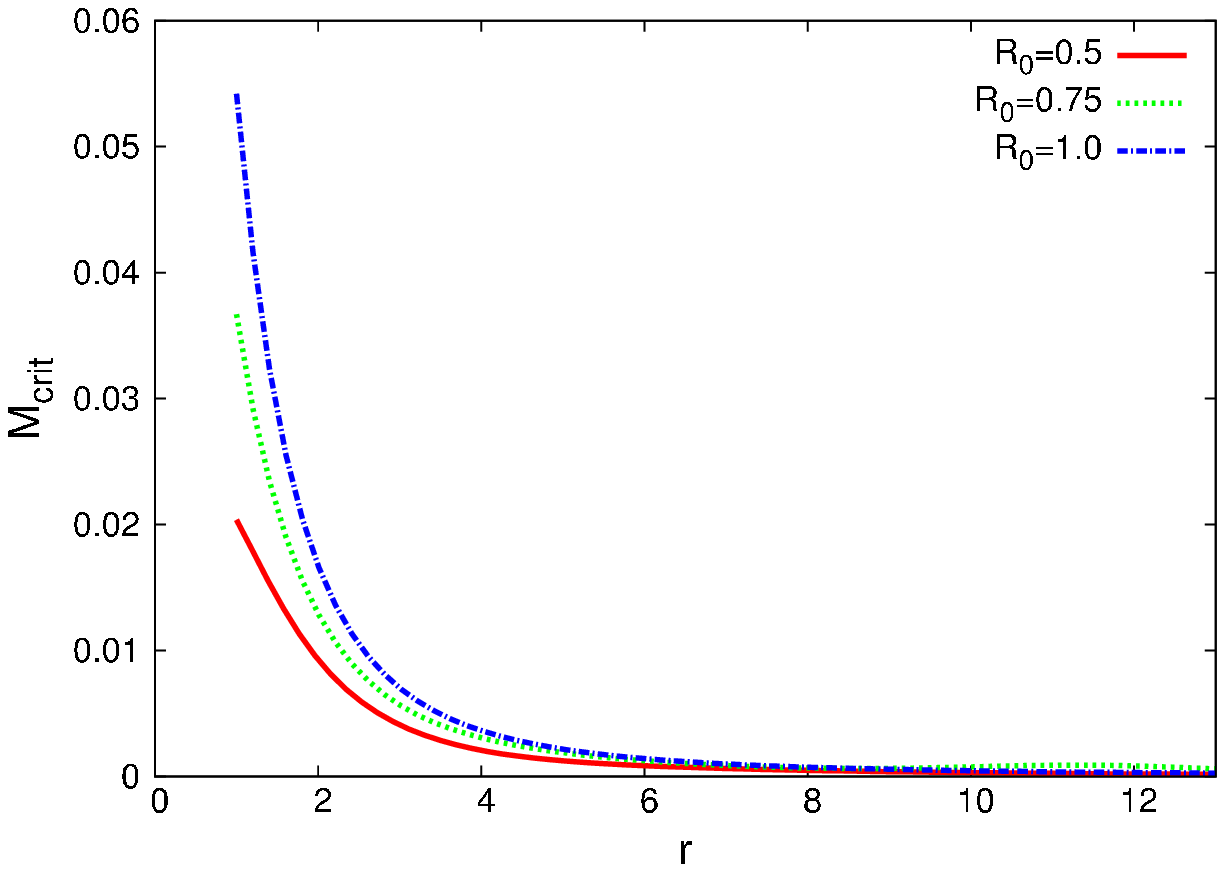}
\caption{Critical mass for the radial geodesics in the JMN background for different values of the radius of the star $R_0$ with $M_0$ = 0.1}
\label{radjmn}
\end{minipage}
\end{figure}
Next, we come to the tidal potentials, computed from Eq.(\ref{phit}). We show this for two examples. 
Figs.(\ref{bsttidpot2}) and (\ref{bsttidpot}) shows the variation of the second order tidal potential along the three axes in the Fermi normal coordinates, 
for $r=3$ and $r=8$. Although the plots have been shown for BST, we have checked that the nature of the tidal potential is similar for other metrics, 
for circular geodesics. The gradient of the plot is a measure of the tidal force. The given plots indicate that the potential is confining along the 
$q^2$ and $q^3$ axes and hence the resulting equilibrium configuration will have major axis along $q^1$. Note that the leading order
tidal potential has a symmetry about the origin along all the axes. This symmetry is generically broken in small amounts when higher order terms 
in the tidal potential are taken into account. We observe  from Figures (\ref{bsttidpot2}) and (\ref{bsttidpot}) that as $r$ is decreased, 
the tidal potential becomes steeper, indicating higher tidal effects. For the Kerr case, we observed that the potential is less steep along the 
$q^1$ and $q^2$ axes for a higher value of $a$ indicating that that the tidal forces weaken as the effect of spin is taken into account. 
\begin{table}
\caption{Comparative Table For Circular Geodesics for $r=8$}
\centering
\begin{tabular}{ |c |c |c | }
\hline\hline
Metric & Parameters & $M_{crit}$ \\
\hline
Schwarzschild & $M=1$ & $2.5 \times 10^{-3}$ \\
JNW & $\nu=0.8$ & $2.8 \times 10^{-3}$  \\
JNW & $\nu=0.6$ & $3.5 \times 10^{-3}$ \\
JNW & $\nu=0.4$ & $6.1 \times 10^{-3}$ \\
BST & $D=10^3$, $\beta=0.1$ & $5.7 \times 10^{-4}$ \\
BST & $D=10^3$, $\beta=0.6$ & $1.9 \times 10^{-3}$ \\ 
BST & $D=10^3$, $\beta=0.8$ & $5.4 \times 10^{-3}$ \\
BST & $D=10^4$, $\beta=0.3$ & $2.8 \times 10^{-3}$ \\
JMN & $M_0=0.2$ & $1.1 \times 10^{-3}$  \\
JMN & $M_0=0.08$ & $4.4 \times 10^{-4}$  \\
JMN & $M_0=0.02$ & $1.1 \times 10^{-4}$  \\
\hline
\end{tabular}
\label{tab1}
\end{table}
It is also interesting to look at the behavior of the density variation of the star near equilibrium.
Figs.(\ref{bstden1}) and (\ref{bstden2}) show the typical variation of the equilibrium density distribution at the Roche limit in the second order tidal 
potential along the $q^1$ axis for different values of $q^2$, where we have chosen the polytropic index to be $0.5$ and $1.5$, respectively, as a 
function of the coordinate $q^1$. In Fig.(\ref{bstden1}), this is shown for $q^2=q^3=0$ (solid red), $q^2=0,q^3=3R/8$ (dashed green), 
$q^2=0,q^3 = R/2$ (dot-dashed blue) and $q^2=0, q^3=5R/8$ (sparse-dotted magenta). In Fig.(\ref{bstden2}), the same values of $q^2$ and $q^3$ are 
used for the solid red and the dashed green curve. In this figure, the dot-dashed blue curve corresponds to $q^2=0, q^3=R/4$.

The density in equilibrium is set to zero at $q^1=0.8$ along the major axis on the grid 
corresponding to $R_0$. As expected, the density along the $q^2$ direction goes to zero faster than in the $q^1$ direction. We observe that the 
density is symmetric with respect to the $q^1=0$ plane. This symmetry is generically broken slightly when higher order terms in the tidal potential 
are taken into consideration. For smaller values of the polytropic index, the pressure, in accordance with the stiffer polytropic equation of state, 
has a stronger density dependence. This is reflected in the fact the equilibrium configuration is more compact with higher surface density for 
$n=0.5$ whereas for $n=1.5$, the density goes to zero smoothly at the surface due to lower pressures. 

Finally, in Figs.(\ref{radjnwn}) and (\ref{radjmn}), we show the critical mass versus radial distance for the JNW and JMN singularities. In 
Fig.(\ref{radjnwn}), we show the variation of the critical mass for different values of the polytropic index, with the solid red, dotted green and
dot-dashed blue curves corresponding to $n=0.5$, $1$ and $1.5$ respectively. In Fig.(\ref{radjmn}), this is shown as a function of the
radius of the star, with the solid red, dashed green and dot-dashed blue lines corresponding to this radius being $0.5$, $0.75$ and $1$,
respectively. 

We now tabulate some numerical data on the various cases that we have discussed. In table (\ref{tab1}), 
we have provided such data for circular geodesics in the Schwarzschild, JNW, BST and JMN metrics, with the mass
of the singularity taken to be unity in all the cases. One can see from the table that at this value of the radial distance, the tidal force can be of
similar orders of magnitude for the Schwarzschild black hole, and the JNW and the BST naked singularity backgrounds. 
\begin{table}[t!]
\caption{Comparative Table For Radial Geodesics for $r=8$, $E=0.01$}
\centering
\begin{tabular}{ |c |c |c | }
\hline\hline
Metric & Parameters & $M_{crit}$ \\
\hline
Schwarzschild & $M=1$ & $1.5 \times 10^{-3}$ \\
JNW & $\nu=0.8$ & $1.9 \times 10^{-3}$  \\
JNW & $\nu=0.6$ & $2.1 \times 10^{-3}$ \\
JNW & $\nu=0.4$ & $4.7 \times 10^{-3}$ \\
BST & $D=10^3$, $\beta=0.1$ & $1.4 \times 10^{-4}$ \\
BST & $D=10^3$, $\beta=0.6$ & $3.4 \times 10^{-3}$ \\
BST & $D=10^3$, $\beta=0.8$ & $3.9 \times 10^{-3}$ \\
BST & $D=10^4$, $\beta=0.3$ & $1.3 \times 10^{-2}$ \\
JMN & $M_0=0.2$ & $5.4 \times 10^{-4}$  \\
JMN & $M_0=0.08$ & $2.7 \times 10^{-4}$  \\
JMN & $M_0=0.02$ & $8.1 \times 10^{-5}$  \\
\hline
\end{tabular}
\label{tab2}
\end{table}
Specifically, tidal forces are seen to be stronger as one decreases the value of $\nu$ in the JNW background. For example, at 
$\nu = 0.4$, the critical mass allowed by this background is about three times that in a same-mass Schwarzschild background. This 
would imply that if a stellar object of higher critical mass than predicted by a Schwarzschild analysis is found, this might indicate that one
needs to look closer at the nature of the central singularity. On the other hand, the JMN background indicates that stellar objects having
masses much smaller (by even two orders of magnitude) than those predicted by a Schwarzschild analysis might exist in stable circular 
orbits. Although direct observational evidence for such objects might be extremely difficult in practise, nonetheless our simple minded
analysis indicates useful bounds on the masses of stellar objects in circular orbits of a given radius.  Of course, we have confined our analysis to 
circular orbits, whereas actual orbits of stellar objects might be highly elliptical. We however expect that qualitative features of our analysis will 
remain unchanged in such situations also, though this requires a more detailed analysis. 

As a consequence of our analysis, we note that for large values of $r$, the differences in the magnitude of the tidal force 
between the Schwarzschild and the JNW naked singularity background increases appreciably. For example, for circular geodesics at $r=12$, we find that
whereas $M_{\rm crit} = 9.8 \times 10^{-4}$ for the Schwarzschild case, it is $5.6 \times 10^{-3}$ for the JNW naked singularity 
with $\nu = 0.2$. For circular geodesics at $r=21$, the Schwarzschild critical mass is $8.6 \times 10^{-3}$, while the JNW background in this
case with $\nu = 0.1$ predicts a critical mass $5.5 \times 10^{-4}$, almost an order of magnitude difference. 
This latter set of numbers translate into interesting realistic ones
as we now illustrate. Take the mass of the central singularity as $1.8 \times 10^3 M_{\odot}$, and $R_0=14~{\rm km}$ (which correspond to
$R_0 = 5\times 10^{-3}$ in natural units). Then, $r=21$ in natural units leads to $r = 5.6\times 10^4~{\rm km}$. The Schwarzschild 
background in this case predicts a critical mass $M_{\rm crit}=1.0M_{\odot}$, while a JNW background with $\nu = 0.1$ predicts
$M_{\rm crit} = 15.7 M_{\odot}$, where we have assumed $n=1$ in the polytropic equation of state of the neutron star. Given a typical neutron 
star mass $\sim 1.5 M_{\odot}$, we see that the neutron star is tidally disrupted for the naked singularity background but not for the Schwarzschild 
background at this radius. 

It is also worthwhile to mention that while the discussion of the previous paragraph focused on the JNW naked singularity, the result assumes
significance given the fact that similar conclusions can be reached for BSTs as well. Indeed, from Fig.(\ref{bstcirc}), we see indications that 
the critical mass increases sharply as the value of the parameter $\beta$ in Eq.(\ref{BSTmetric}) is increased. It is thus expected that 
appropriately tuning the parameters of a BST, one can in principle obtain values of the critical mass substantially larger than what is obtained
in Schwarzschild backgrounds. On the other hand, as Table (\ref{tab1}) indicates,  JMN backgrounds typically show lower tidal effects compared
to the Schwarzschild cases. 

Table (\ref{tab2}) summarizes the results for the radial geodesics, which have to be understood along with the limitations discussed in subsection 2.5. 
We see that these are qualitatively similar to those for the circular geodesics. In this table, we have considered a small energy of the star. Increase in this 
leads to a higher difference in magnitude for the critical mass for the naked singularity background compared to the Schwarzschild case. 
However, the assumption that the star has travelled a distance that is small compared to the scale over which the tidal forces change
will not be valid for high energies of the star, and thus higher values of energy will probably not be very trustable in our framework. 

Before we end, let us briefly comment on the nature of tidal forces in the interior Schwarzschild solution of Eq.(\ref{IntSchmetric}). We considered
radial geodesics in this geometry, with energy $E=0.01$. Here, we have taken a matching radius $r=20$. Our results in this case suggest that the critical 
mass reaches a saturation value as the radial distance of the stellar object decreases. However, we were able to achieve good numerical convergence 
in this case only for a limited number of parameter values. This case is of interest as this exemplifies motion in a diffused matter background by a stellar
object (which is assumed not to back react on the background) and merits further understanding, which we defer for a later study. 

\section{Discussions and Conclusions}

In this paper, we have examined the nature of tidal forces in a class of non-rotating naked singularity backgrounds on stellar objects
in radial and circular geodesic motion. The purpose of this work was
to understand theoretical differences in the nature of these forces, keeping in mind observational aspects. Broadly, this paper establishes the 
magnitudes of these forces for three different naked singularity backgrounds. Our main conclusion here is that tidal forces can be significantly 
different for these cases when compared to the Schwarzschild background. If in experiments can indicate numerical bounds on the masses of objects
that can be tidally disrupted by a singularity, our results might be effective in modelling such situations, to indicate the nature of the underlying
singularity. 

In this paper, we have confined ourselves to a second order expansion of the tidal potential in Fermi normal coordinates. The analysis of \cite{ishii},
which is based on a fourth order expansion of the potential is numerically more accurate. However, in this paper our main focus was comparing
the magnitudes of the tidal force in different backgrounds. Inclusion of higher order corrections, although important, will not qualitatively affect
the results of the present analysis. 

Also, our analysis is simple minded, has a number of constraints which we have extensively discussed. Nevertheless, we believe that the 
present analysis complements the work of \cite{ishii} and sets the stage for a deeper question, namely if tidal disruptions can be an effective indicator of 
the central singularity in galaxies or galaxy clusters. As an immediate future application, we hope to build upon the results of this paper to 
understand rates of tidal disruption events, while taking into account the fluid nature of the star. This work is in progress. Further, as is well known,
a realistic neutron star has a superfluid character, and its fluid dynamics might be different from what is assumed in this work. How this will affect 
tidal forces on neutron stars might be an important issue to understand. \\

\noindent
{\bf Note added :} A preliminary version of the FORTRAN code used for the computations in this paper is available upon request.

\end{document}